# First-principles study of structural stability, dynamical and mechanical properties of Li$_2$FeSiPO$_4$ polymorphs

P. Vajeeston,[*] and H. Fjellvåg

*Center for Materials Sciences and Nanotechnology, Department of Chemistry, University of Oslo, P.O. Box 1033 Blindern, N-0315 Oslo, Norway*

**Abstract**

Li$_2$FeSiO$_4$ is an important alternative cathode for next generation Li-ion batteries due to its high theoretical capacity (330 mA h/g). However, its development has faced significant challenges arising from structural complexity and poor ionic conductivity. In the present work, the relative stability, electronic structure, thermodynamics, and mechanical properties of potential cathode material Li$_2$FeSiO$_4$ and its polymorphs have been studied by state-of-the-art density-functional calculations. Among the 11 structural arrangements considered for the structural optimization calculations, the experimentally known monoclinic $P2_1$ modification is found to be the ground state structure. The application of pressure originates a sequence of phase transitions according to ***P2$_1$ → Pmn2$_1$ → I222***, and the estimated values of the critical pressure are found to be 0.38 and 1.93 GPa. The electronic structures reveal that all the considered polymorphs have a non-metallic character, with band gap values varying between 3.0 and 3.2 eV. The energy differences between different polymorphs are small, and most of these structures are dynamically stable. On the other hand, the calculation of single crystal elastic constants reveals that only few Li$_2$FeSiO$_4$ polymorphs are mechanically stable. At room temperature, the diffusion coefficient calculations of Li$_2$FeSiO$_4$ in different polymorphs reveal that Li-ion conductivity of this material is destitute.

[*] Electronic address: ponniahv@kjemi.uio.no; http://www.folk.uio.no/ponniahv

**Introduction**

The research on polyanion cathodes for lithium-ion batteries has continued to gain momentum since Padhi *et al*. reported the electrochemical properties of LiFePO$_4$ in 1997 [1]. The interest in polyanion cathodes comes from added safety and higher voltage values in comparison to the oxide analogues with the same $M^{2+/3+}$ redox couples. The increased safety and higher voltage values have been attributed to strong covalent bonding within the polyanion units. Over the years, these inherent characteristics of polyanion cathodes have promoted the investigation of several series of polyanion compounds for use in lithium-ion batteries. For example: Li$_2$*M*SiO$_4$ silicates, Li$_2$*M*P$_2$O$_7$ pyrophosphates, and Li*M*BO$_3$ borates (*M* = Mn, Fe, Co, and Ni). Each of these compounds possesses additional favourable characteristics as cathode materials. The borates contain the lightest of the polyanion units (BO$_3$) and, therefore, have a higher theoretical capacity (~200 mA h g$^{-1}$) than LiFePO$_4$ (~170 mA h g$^{-1}$). The pyrophosphates and silicates offer the appealing possibility of extracting/inserting two lithium ions per transition metal ion in the material, further increasing the theoretical capacity, respectively, to ~220 and 330 mA h g$^{-1}$. Additionally, silicon is one of the most abundant elements on earth's crust, offering a reduction in cost for the silicates. Li$_2$FeSiO$_4$ (hereafter referred as LFS) is an attractive member of Li$_2$*M*SiO$_4$ group, which is built on inexpensive and abundant raw materials [2,3]. In recent works, small particles of LFS and proper electrode engineering have yielded electrodes with good cycling stability close to the theoretical capacity [4]. Experimental and theoretical studies on LFS have found stable structures where the cations are located in the tetrahedral interstitials of a nearly hexagonally close-packed framework of oxygen atoms [5-9]. Depending on the crystal structure, the Li ions in transition metal silicates can either be arranged in layers [10], along lines, or in a three-dimensional network. Hence, the ionic conductivity might be strongly anisotropic [11,12]. The most known polymorphs are *Pmn*2$_1$ (orthorhombic at low temperature), *P*2$_1$(monoclinic at 700$^o$C), and *Pmnb* (orthorhombic at 900$^o$C) [13]. However, the monoclinic space group *P*2$_1$ is found to be the most stable polymorph by means of first-principles calculations [10,13-15]. The stability of these polymorphs of LFS plays an important role in understanding the multi-electron process of LFS. However, to the best of our knowledge, the study of dynamical and mechanical stability of LFS is still missing to date. In the present study, we have investigated the structural phase stability, electronic, mechanical, and lattice dynamical properties of the of LFS polymorphs.

**Computational details**

Total energies have been calculated by the projected-augmented plane-wave (PAW) implementation of the Vienna *ab initio* simulation package (VASP) [16,17]. All these calculations were made using the Perdew, Burke, and Ernzerhof (PBE) [18] exchange-correlation functional with the Hubbard parameter correction (GGA+$U$), following the rotationally invariant form [19,20]. An effective $U$ values of 5eV (with $J$=1 eV) was used for the Fe-$d$ states. Ground-state geometries were determined by minimizing both the stresses and the Hellman-Feynman forces using the conjugate-gradient algorithm with force convergence threshold of $10^{-3}$eV Å$^{-1}$. Brillouin zone integration was performed with a Gaussian broadening of 0.1 eV during all relaxations. From various sets of calculations, it was found that 512 **k** points in the Brillouin zone for the structure with a 600 eV plane-wave cut-off are sufficient to ensure optimum accuracy in the computed results. The **k**-points mesh was generated using the Monkhorst-Pack method with a grid size of 8×8×8 for structural optimization. A similar density of **k**-points and energy cut-off were used to estimate total energy as a function of volume for all the structures considered in the present study. Iterative relaxation of atomic positions was stopped when the change in total energy between successive steps was smaller than 1meV/cell. From our total-energy calculation for the LFS-$Pmn2_1$ phase, we have found that the antiferromagnetic (AFM) and ferromagnetic (FM) states are lower in energy compared to the paramagnetic (PM) state. The energy difference between the AFM and FM states is found to be 2 meV/cell. (i.e., comparable with the convergence threshold), moreover, both states have a similar cell volume (see Figure 1). Hence we have considered only the FM states in the rest of this work.

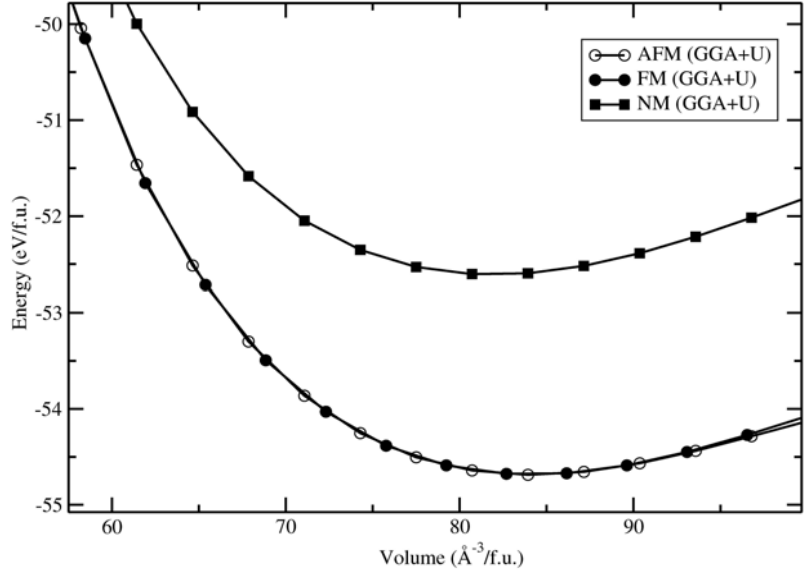

**Figure 1** Calculated unit cell volume vs. total energy (per formula unit; f.u.) for $Li_2FeSiO_4$ in *Pmn21* symmetry with different magnetic configuration (anti ferro magnetic-AFM, ferro magnetic–FM, and nonmagnetic-NM ordering).

The calculated total energy as a function of volume has been fitted to the universal equation of state (EOS) [21]. The transition pressures are calculated from the pressure vs. Gibbs free energy curves. The Gibbs free energy ($G = U+PV-TS$ where $T = 0$; $G$ = total energy + pressure×volume) is calculated in the following way: the volume versus total energy curves calculated for two data sets were fitted to the universal EOS function. The pressure is defined as $P = (B_0/B_0^/) \times [(v_e/v)^{B0'} - 1]$. The relation can be inverted to obtain the volume as $(v) = v_e / [ (1 + (B_0^// B_0 \times p)^{1/B0'}]$ where $v_e$, $B_0$, and $B_0^/$ refers to the equilibrium volume, the bulk modulus, and its derivative with respect to pressure, respectively. The inverse is then calculated using the bisection method. From the scan over the pressures, the corresponding difference in the enthalpy between the two data sets was calculated.

A frozen phonon calculation was performed on the supercells using the phonopy program to obtain the phonon dispersion curve and density of states [22]. An atomic displacement of 0.0075 Å was used, with a symmetry consideration, to obtain the force constants for the phonon calculations. The displacements in opposite directions along all axes were incorporated in the calculations to improve the overall precision. The force calculations were made using the VASP code with the supercell approach (with GGA+*U* correction) and the resulting data were imported into the Phonopy program. The dynamical matrices were calculated from the force constants, and phonon DOS curves were computed using the Monkhorst-Pack scheme.[23] The thermal properties, such as heat capacity, free energy and entropy, were obtained using the calculated phonon density of states (DOS).

The Li diffusion barrier height of the different polymorphs are investigated with the cNEB method using supercell approach [24,25]. A large supercell (dimensions listed in Table 2) was used to ensure that the atoms are separated from their periodic images, providing a more accurate result for the activation barrier in the diluted limit. To determine the minimum energy path (MEP) through the climbing Nudged Elastic Band (cNEB) method, six replicas of the system were created, in each of which the diffusing Li atom was moved by equidistant steps to intermediate positions between the initial and final states, as obtained by linear interpolation of the path. A 1×2×1 supercell (for $P2_1$; the supercell sizes for the other phases are reported in Table 2) was used to ensure that the atoms are separated from their periodic image, providing a more accurate answer for the activation barrier in the diluted limit.

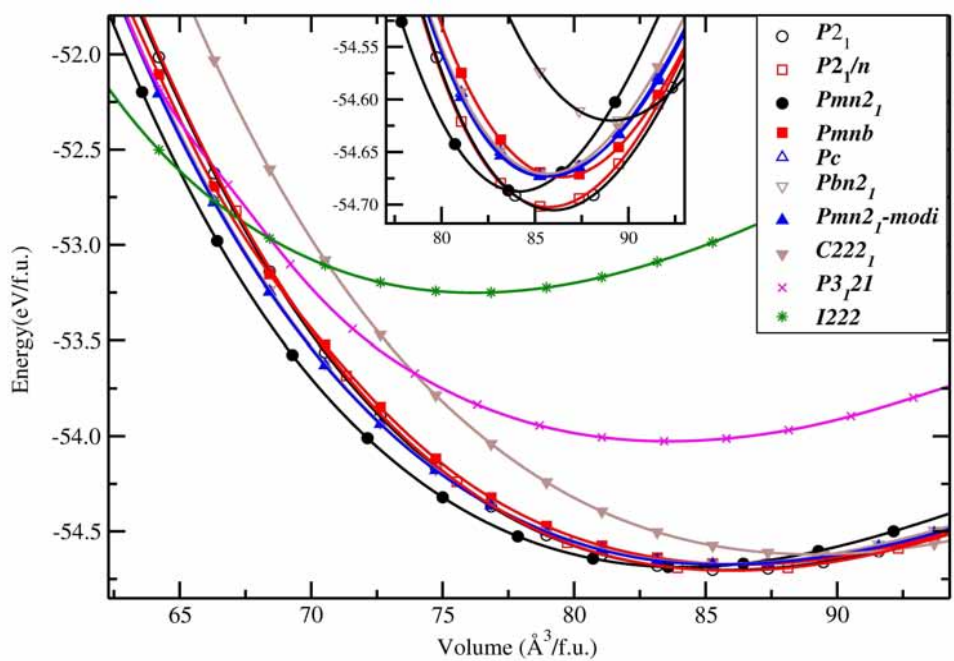

**Figure 2** Calculated unit cell volume vs. total energy (per formula unit; f.u.) for $Li_2FeSiO_4$ in actual and possible structural arrangements (structure types being labeled on the illustration). The magnified version around the low energy part of the cell volume vs. total energy is shown as an insert.

**Result and Discussions:**

**Structure models Considered:**

The crystal structure of LFS is quite ambiguous due to its rich polymorphism and hence to the difficulties encountered in obtaining single phase samples. The early studies by Tarte and Cahay [26], and Nyten *et al.* [26] shows that the structure of LFS is isostructural with β-Li$_3$PO$_4$, that is, crystallizing in the orthorhombic structure with space group *Pmn*2$_1$. As noted by Quoirin *et al.*[27], the indexation given in ref. [28] was highly questionable, and it was found that LFS under goes a series of complex phase transformations. The *Cmma* space group with *a* = 10.66 Å, *b* = 12.54 Å, *c* = 5.02 Å was proposed for LFS annealed at 800°C. Nishimura *et al.* [28] determined the crystal structure of LFS (synthesized from a ceramic-type route at 800°C) in monoclinic symmetry with *a* = 8.23 Å, *b* = 5.02 Å, *c* = 8.23 Å, β = 99.20° [9], which was confirmed by Boulineau *et al.* [28] and Sirisopanaporn *et al.* who discoverd and calculated the crystal structure of a new metastable polymorph, obtained by rapid quenching at ambient temperature from 900°C.[14] In general, the structural models describing LFS are derived from Li$_3$PO$_4$-based structures, in which one-half of the tetrahedral sites, generated by a distorted hexagonal close packing of oxygen atoms, are occupied by cations. Li$_3$PO$_4$ itself crystallizes in two main groups of polymorphs (named as β and γ), which differ in their respective orientations of filled tetrahedral: all T$^+$ (oriented upward) in the low-temperature β form, T$^+$ and T$^-$ (oriented downward) for the high-temperature γ form [29]. The structures of Li$_2$*M*SiO$_4$ analogues (*M* = Zn, Mn, Mg, and Co) have been reported to adopt "simple" β-type or γ-type structures or their distorted derivatives [7,30-33]. The relative stability and electrochemical properties of various LFS polymorphs were very recently investigated from first-principles calculations [10,34]. Quite recently the surface structures and energetics of the *Pmn*2$_1$ polymorph of the LFS were studied using DFT [35].

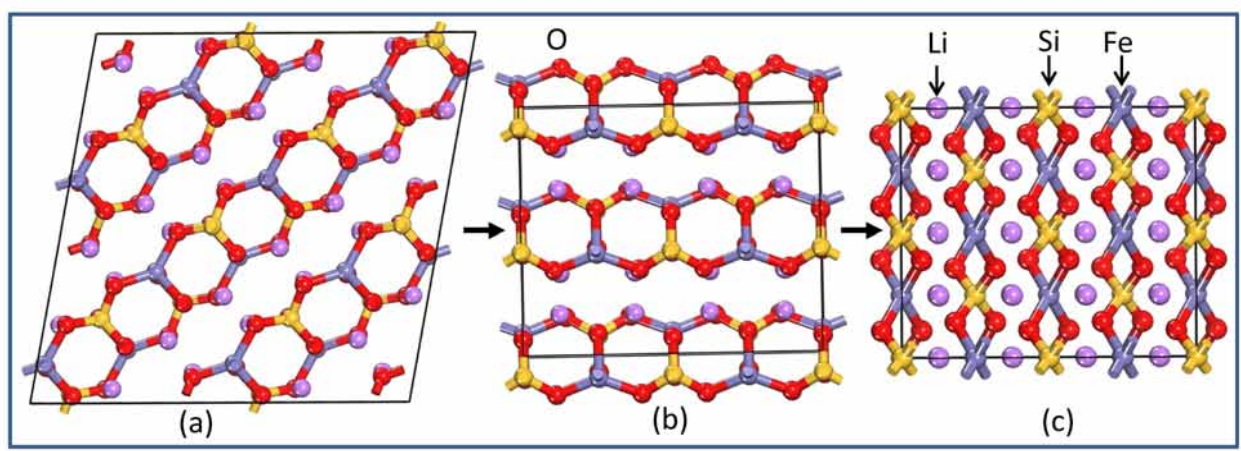

**Figure 3** Theoretically predicted low energy crystal structures for Li$_2$FeSiO$_4$: (a) *P*2$_1$ structure and (b) *Pmn*2$_1$ structure at equilibrium volume and (c) *I*222 structure at 1.93 GPa at T=0K. The legends for the different kinds of atoms are given in the illustration.

In general, the equilibrium crystal structures predicted based on first principles calculations using the structural inputs from Inorganic crystal structure database (ICSD) [36] mostly agree well with experimental structures. In our long experience (see Ref. [37-39] ) on predicting structural properties of hydrides and oxides we found that the ICSD/guess-structure approach is more reliable when a number of existing structural information is used as a starting point. The reliability of the calculation depends upon the number of input structures considered in the calculations. Though it is a tedious process to select input structures from the 486 entries for the $ABC_2X_4$ composition in the ICSD database, which also involves extensive computations, several compounds/phases have the same structure type and some cases have only small variation in the positional parameters (only for certain atoms). Even though we used different positional parameters, these structures converted mostly to the similar type of structural arrangement during the full geometry optimization and hence these possibilities are omitted. For our theoretical simulations, we have chosen 11 structure types from the $ABC_2X_4$. The involved structure types are (space group and space group number are given in the parenthesis): Li$_2$FeSiO$_4$($P2_1$; 4), Li$_2$BeSiO$_4$($Pc$; 7), Li$_2$FeSiO$_4$($P2_1/c$; 14), Li$_2$BeSiO$_4$($C222_1$; 20), Li$_2$FeSiO$_4$($Pmn2_1$; 31), Li$_2$CoSiO$_4$($Pna2_1$; 33), Li$_2$FeSiO$_4$($Pnma$; 62), Li$_2$CaSiO$_4$($I$-$42m$; 121), EuLi$_2$SiO$_4$($P3_121$; 152), Li$_2$BaSiO$_4$($P6_3cm$; 185), and Li$_2$MnSiO$_4$-modified-$Pmn2_1$. Among the considered structures for our structural optimization, the calculated total energy at the equilibrium volume for the $P2_1$ atomic arrangements occur at the lowest total energy (see Figure 2). The calculated positional and lattice parameters (see Table 1) are found to be in good agreement with experimental findings [9] and with the other theoretical calculations [10,34]. It consists of a lattice built up from infinite conjugated layers of composite SiFeO$_4$ linked through the LiO$_4$ tetrahedra, with each Li, Fe, Si located in the centre of the tetrahedron formed by four oxygen atoms. Furthermore, the Li$^+$ ions are occupying tetrahedral sites between the FeO$_4$–SiO$_4$ where the tetrahedra alternately point in opposite directions. It should be noted that this phase is stabilized when the particle size is reduced [40] and, according to experimental findings, it is a thermodynamically less stable phase [9,41]. The next energetically favourable phase is orthorhombic $Pmn2_1$. In this structure, chains of LiO$_4$ tetrahedra run along the *a* direction, parallel to the chains of alternating FeO$_4$ and SiO$_4$ tetrahedra (see Fig. 3). The energy difference between this phase with $P2_1$ at the equilibrium volume is only ca. 4.6 meV/f.u (see Figure 2). The calculated structural parameters are found to be in good agreement (see Table 1) with the recent experimental finding [9] and with other DFT studies [10,34]. It is interesting to note that the energy difference between the $P2_1/c$, $Pnma$, $Pc$, $Pna2_1$ and modified-$Pmn2_1$ is also very small, and hence, one can easily modify one polymorph into another by application of temperature or pressure; this explains the difficulties to control the synthesis of single phase samples of LFS polymorphs, also related to very similar electrochemical properties (voltage, volume variation, and electronic structure) [10,42]. One should also remember that the calculated results are valid only for defect-free ideal materials at low temperatures. However, the experimental findings show that, depending upon the synthesis route/conditions, one can stabilize different polymorphs of LFS and it is therefore difficult to get phase-pure materials from most synthesis processes [9,14,41].

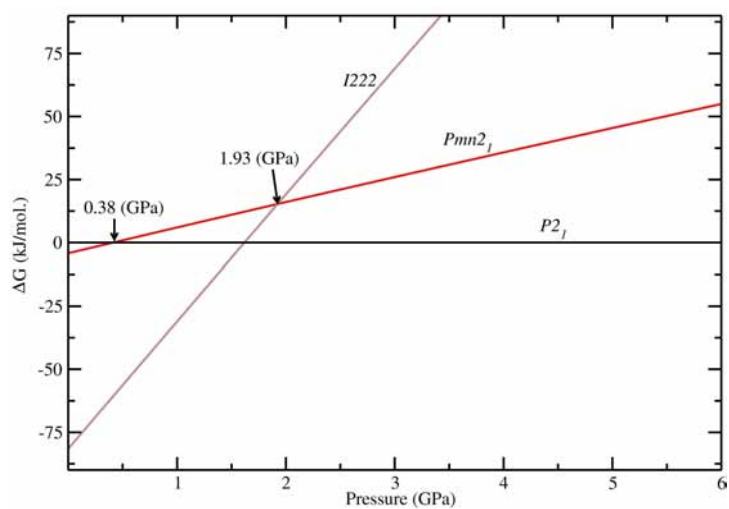

**Figure 4** Calculated stability diagram for $Li_2FeSiO_4$ phases (difference in Gibbs free energy ΔG) related to low energy $P2_1$ structure at T=0K. The transition points are marked by arrow with corresponding pressure.

As discussed above, $P2_1$ is the ground-state structure, and it transforms into $Pmn2_1$ modification at 0.38 GPa (see Figures 2 and 3). Further application of pressure on this $Pmn2_1$ modification transforms it into $I222$ modification at 1.9 GPa (see Fig.4). It should be noted that the $I222$ structure is very close related to the I-42m (space group 121) and both modifications have similar energies (the variation is only 0.2 meV/f.u.). The pressure induced $Pmn2_1$-to-$I222$ transition involves reconstructive rearrangements of the Li, Fe, Si, and O lattices with breaking and reconstruction of bonds. Usually, the application of pressure reduces the covalency in solids and makes the valence electrons more diffuse than at ambient condition. All these three polymorphs $P2_1$, $Pmn2_1$ and $I222$ are stabilized in layered structure type with a similar type of atomic arrangement. The main difference between the three polymorphs are both $P2_1$, $Pmn2_1$ polymorphs are having almost similar interlayer distances while in the $I222$ phase the interlayer distance become very narrow and the $SiO_4$ and $FeO_4$ tetrahedra become stretched along [011] (see Figure 3).

The electronic up-spin and down-spin band structure calculated at the equilibrium volumes for all polymorphs considered in this study for LFS are displayed in Figure S1-S8 (see supporting information). All modifications have finite energy gap ($E_g$; vary from 3.06 to 3.24 eV) between the valence band (VB) and the conduction band (CB), and hence, they have a non-metallic character. The magnitude of the $E_g$ and band positions suggests that the LFS polymorphs are direct wide bandgap semiconductors. It is well known that the bandgap ($E_g$) values of solids obtained from usual DFT calculations are systematically underestimated due to the discontinuity in the exchange-correlation potential. Thus the calculated $E_g$ values are commonly 30-50% smaller than those measured experimentally. Hence, LFS polymorphs are likely to have a much larger bandgap value than the one identified in this study.

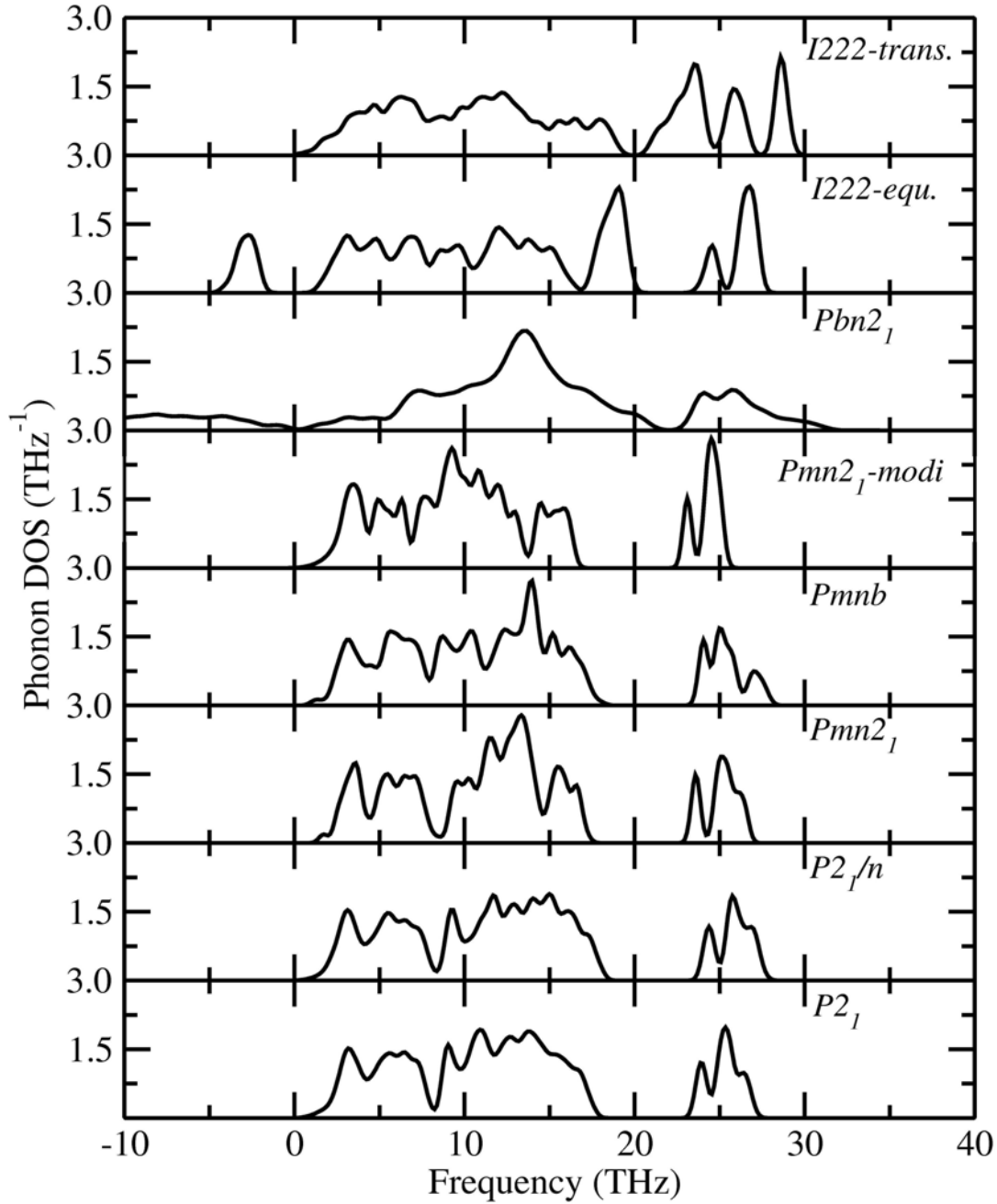

**Figure 5** Calculated total phonon density of states for $Li_2FeSiO_4$ in different modifications. The modifications are noted in the corresponding panel.

The total phonon density of states (PDOS) is calculated at the equilibrium volumes for different polymorphs of LFS. For the $I222$ modifications, the PDOS are calculated both at equilibrium and at the phase transition point. The calculated PDOS of LFS polymorphs are displayed in Figure 5. For all these polymorphs, no imaginary frequency was observed (except for Pbn21 and $I222$ phase), indicating that all the structures (except Pbn21 and $I222$) are stable or at least dynamically stable at ambient conditions. In the $I222$ phase the ambient condition phonon, soft modes disappear at pressures above 1.9 GPa (see Figure 5). Hence, the predicted $I222$ phase is stable above the critical pressure. All the polymorphs of LFS (except Pbn21) including the high-pressure phase have a similar phonon density of states. Hence, we have displayed in Figure 6 only the partial phonon DOS for $Pmn2_1$ polymorphs. In the $Pmn2_1$ phase, the partial phonon DOS is plotted along three directions $x$, $y$, and $z$. For Fe, Si, and Li atoms, the vibrational modes along the $x$, $y$, and $z$ directions are almost identical. On the other hand, for the O atoms, the vibrational modes along the $x$, $y$, and $z$ directions are considerably different at the high-frequency region (see Figure 6).

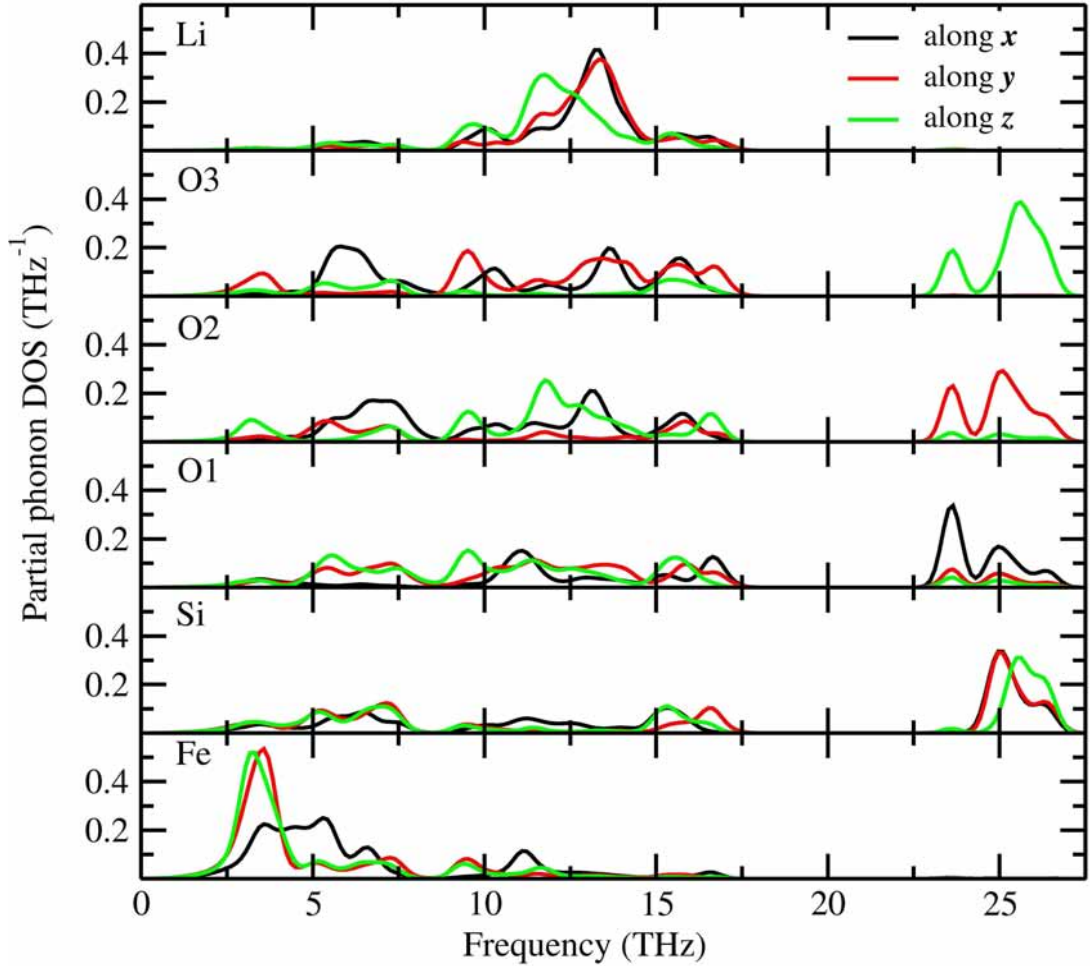

**Figure 6** Calculated site projected phonon density of states for $Li_2FeSiO_4$ in $Pmn2_1$ modification. The corresponding sites are noted in the panel and the phonon mode along different directions are marked with different color (black-along $x$; red-along $y$; and green-along $z$).

In addition to dynamical stability, some important thermodynamic properties of LFS polymorphs (such as, free energy, entropy, and lattice heat capacity; see Figure S9) at zero K, is calculated. Our calculated temperature-dependent lattice specific capacities for the different polymorphs of the LFS have similar values except for the high-pressure polymorph (see Figure 7). This finding clearly demonstrates that all these polymorphs have similar thermodynamic properties. The specific heat coefficient increases rapidly below 600K. Above this temperature the slope becomes gentler.

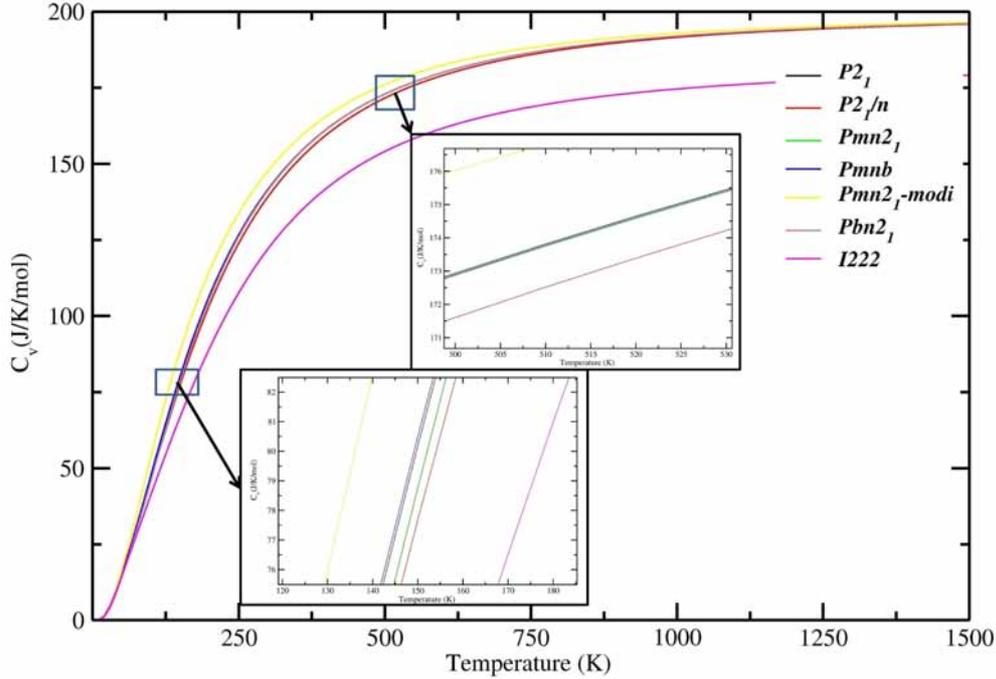

**Figure 7** Calculated lattice heat capacity verses temperature plot for $Li_2FeSiO_4$ polymorphs. The magnified version around the 100K and 500K part of the heat capacities curves are shown as an insert.

**Single crystal elastic constants and mechanical stability**

To understand the mechanical stability of the considered phases, we have computed the single-crystal elastic constants using the finite strain technique. The elastic constants of a material describe its response to an applied strain or, conversely, the stress required to maintain a given deformation. Both stress and strain have three tensile and three shear components. The linear elastic constants of a crystal can therefore be described using a 6×6 symmetric matrix, having 27 different components, 21 of which are independent. However, any symmetry present in the structure may reduce this number. Properties such as the bulk modulus (response to an isotropic compression), Poisson coefficient and Lame constants can be computed from the $C_{ij}$ matrix. Methods to determine the elastic constants from first principles usually involve setting either the *stress* or the *strain* to a finite value, re-optimizing any free parameters and calculating the other property (the strain or stress, respectively). Applying a given homogeneous deformation (strain) and calculating the resulting stress requires far less computational effort, since the unit cell is fixed and does not require optimization. The finite strain technique has been successfully used to study the elastic properties of a range of materials including simple metals (such as Al) [43], super hard nitrides [44], borides [45,46], oxides [47], silicates [48], and semiconductors [49]. The results of these studies show that the accuracy of DFT elastic constants is typically within 10% of the experiment. This allows us to predict elastic constants for LFS polymorphs.

For an orthorhombic crystal, the independent elastic stiffness tensor reduces to nine components $C_{11}$, $C_{22}$, $C_{33}$, $C_{44}$, $C_{55}$, $C_{66}$, $C_{12}$, $C_{13}$ and $C_{23}$ in the Voigt notation[48]. The well-known Born stability criteria[50] for an orthorhombic system are

$$B1 = C_{11} + C_{22} + C_{33} + 2(C_{12} + C_{13} + C_{23}) > 0, \qquad (1)$$
$$B2 = C_{11} + C_{22} - 2C_{12} > 0, \qquad (2)$$
$$B3 = C_{11} + C_{33} - 2C_{13} > 0, \text{ and} \qquad (3)$$
$$B4 = C_{22} + C_{33} - 2C_{23} > 0. \qquad (4)$$

All the nine calculated independent single crystalline elastic stiffness constants for LFS in *Pmn2$_1$, Pna2$_1$, Pmna*, and *I222* are given in Table 4. The computed *B1*, *B2, B3,* and *B4* values for the LFS in *Pmn2$_1$* is 811, 237, 236, 177 GPa, in *Pna2$_1$* is 657, 161, 145, 186 GPa; in *Pmna* is 709, 205, 222, 139 GPa, and in *I222* is 1032, -92, -89, 373 GPa. All the four conditions for mechanical stability given in Equations (1-4) are simultaneously satisfied for *Pmn2$_1$, Pna2$_1$,* and *Pmna* and this clearly indicates that these orthorhombic phases are mechanically stable. It should be noted that *I222* polymorphs are unstable at equilibrium conditions and equations *B2* (-92) and *B3* (-89) are correspondingly not satisfied. On the other hand, above the phase transition point, the calculated *Cij* values simultaneously satisfy the equations *B1*(1062), *B2*(18)*, B3*(15)*,* and *B4*(360) and this phase is stable only at high pressure. This finding is consistent with the phonon calculations.

The mechanical stability criteria for the monoclinic phase are given by [51]
$$C_{11} > 0, C_{22} > 0, C_{33} > 0, C_{44} > 0, C_{55} > 0, C_{66} > 0, \qquad (5)$$
$$[C_{11} + C_{22} + C_{33} + 2(C_{12} + C_{13} + C_{23})] > 0, \qquad (6)$$
$$(C_{35} \cdot C_{55} - C^2_{35}) > 0, (C_{44} \cdot C_{66} - C^2_{46}) > 0, (C_{22} + C_{33} - 2C_{23}) > 0, \qquad (7)$$
$$[C_{22}(C_{33} \cdot C_{55} - C^2_{35}) + 2C_{23} \cdot (C_{25} \cdot C_{35} - C^2_{23} \cdot C_{55} - C^2_{25} \cdot C_{33})] > 0, \qquad (8)$$
$$\{2[C_{15} \cdot C_{25}(C_{33} \cdot C_{12} - C_{13} \cdot C_{23}) + C_{15} \cdot C_{35}(C_{22} \cdot C_{13} - C_{12} \cdot C_{23})$$

$$+C_{25}.C_{35}(C_{11}.C_{23} - C_{12}.C_{13})] - [C^2_{15}(C_{22}.C_{33} - C^2_{23}) + C^2_{25}(C_{11}.C_{33} - C^2_{13}) + C^2_{35}(C_{11}.C_{22} - C^2_{12})$$
$$+ C_{55}(C_{11}.C_{22}.C_{33} - C_{11}.C^2_{23} - C_{22}.C^2_{13} - C_{33}.C^2_{12} + 2C_{12}.C_{13}.C_{23})]\} > 0, \qquad (9)$$

The computed independent single crystalline elastic stiffness constants for LFS in $P2_1$, $P2_1/c$, $Pc$, and $Pmn2_1$-mod ($Pc$) at their equilibrium volume are shown in Table 4. In $Pc$ monoclinic polymorph the largest component is $C_{11}$, corresponding to the in-plane strain, and the second largest component, $C_{33}$ is just a few tens of GPa smaller than it. On the other hand, in $P2_1$ and $P2_1/c$ the largest component is $C_{11}$. It is also evident that there is a large degree of elastic anisotropy among the three principal directions due to $C_{11} \neq C_{22} \neq C_{33}$. The elastic constants $C_{15}$, $C_{23}$, $C_{35}$ (for all monoclinic polymorphs) and $C_{46}$ (except for $Pc^*$) become negative and they are very sensitive to the relaxed structural parameters. All the three conditions given in Equations (7-9) are not simultaneously satisfied, and this clearly indicates that all these polymorphs are mechanically unstable phases. This might be the one of the possible reasons why the energetically favourable structure $P2_1$ is a metastable phase in the experimental findings, and different polymorphs coexist during the charge/discharge cycles. The bulk modulus $B$, shear modulus $G$, Young's modulus $E$ and Poisson's ratio $v$ can be assessed from these elastic stiffness moduli through the Voigt ($V$), Reuss ($R$) and Hill ($H$) approximations [52], and the $V$ and $R$ approximations usually give the upper and lower bounds, respectively, of these parameters as indicated in Table 4.

    Like the elastic constant tensor, the bulk and shear moduli contain information regarding the hardness of a material on various types of deformation. Properties such as bulk moduli, shear moduli, Young's moduli and Poisson's ratio can be computed from the values of elastic constants and the calculated values are tabulated in Table 3. All these polymorphs are having almost similar Young's and shear modulus in $x$, $y$ and $z$-direction. The compressibility value of these polymorphs suggested that these polymorphs of LFS are very soft materials. A parameter B/G is also introduced, in which B indicates the bulk modulus and G represent the shear modulus. The bulk and shear moduli are calculated from the Voigt–Reuss–Hill approximations.[53-55] The high (low) B/G value is associated with ductility (brittleness) and the critical value which separates ductile and brittle materials is 1.75 [56]. The calculated B/G values of LFS polymorphs are larger than 1.75, implying the ductile characteristics of materials and the stable cycle performance. It is consistent with the electrochemical measurement that about 98.3% of the discharge capacity of LFS can be retained after 80 charge/discharge cycles [57].

**Diffusion coefficient**

To identify the diffusion pathways and the activation energy for the Li$^+$ transport of the different polymorphs of LiFeSiO$_4$, we have calculated the diffusion constant D using the equation

$$D = d^2 \nu_0 \exp(-E_a/k_B T)$$

where $d$ is the hopping distance, $E_a$ the activation energy, $k_B$ the Boltzmann constant, T the temperature and $\nu_0$ the attempt frequency (assumed as $10^{13}$ Hz.) [58]. For the chosen structure, we have calculated two main Li migration paths with the Li-Li hop: pathway A, consisting of the linear diffusion of the lithium ions along a particular direction (for example in *Pmn21* symmetry the Li hop along *a*-axis) and pathway B, consisting of lithium ions diffusion following a zig-zag trajectory(for example in *Pmn21* symmetry the Li hop along *c*-axis). The Li activation energy ($E_a$) and estimated diffusion coefficients of various polymorphs of LFS along with other theoretical reports are summarized in Table 3. Our calculated activation barriers are in good agreement with other theoretical works [41,42,59-63], with the only exception of the HP phase. The calculated energy barrier for the low energy structure *Pmn21* symmetry is 0.86 and 0.98 eV for pathways A and B respectively. In general, it can be seen in Table 3 that the activation energy ranges from 0.6 eV to 1.64 eV for the Li-ion diffusion in different polymorphs of LFS. The minimum value is observed for P21/n-cycl, and maximum one in the newly identified high-pressure polymorph I222. However, the values listed in Table 3 are much higher than that of ca. 0.30 eV typically reported for the LiFePO$_4$ olivine cathode [64], which directly reflects on the poor ionic conductivity of LFS. Moreover, compared with the other LFS polymorphs, the P21/n-cycl structure has relatively greater activation energy, indicating better Li diffusion in it. This should be ascribed to the more opened 3D-framework structure of P21/n-cycl than the other LFS polymorphs. On the other hand in the high-pressure polymorph I222 the interlayer distances are very narrow, resulting in a larger value for the activation energy.

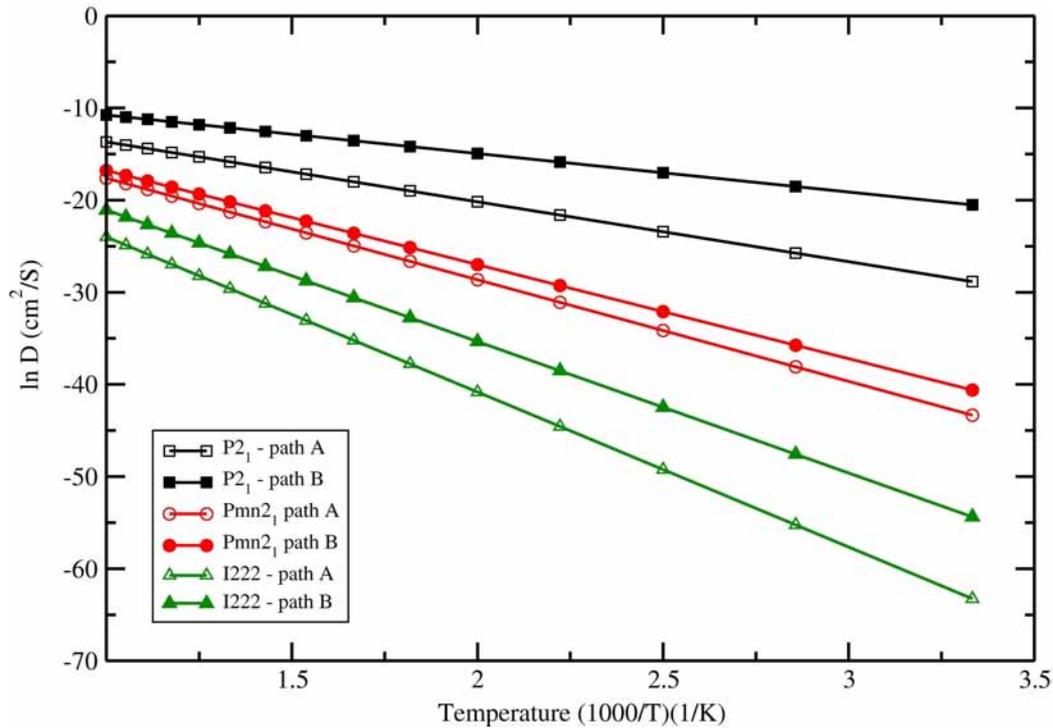

Figure 8 Natural logarithm of the diffusion coefficient against 1/T for $Li_2FeSiO_4$ in different polymorphs with different possible pathways.

Comparing both paths in the *Pmn21* symmetry, it can be noticed that Pathway B would be the most probable one. Figure 8 summarizes the obtained results for the natural logarithm of the diffusion coefficient in the low energy structures of LFS as a function of 1/T. Note that the activation barrier is proportional to the slope of each straight line. In all these three polymorphs the diffusion coefficient follows the trend P21< *Pmn21*< I222. The diffusion coefficient calculated in LFS polymorphs at room temperature, ranging from $10^{-52}$ cm$^2$/s up to $10^{-17}$ cm$^2$/s (lower value in I222 phase and higher value in P21 phase). On the other hand diffusion coefficient in currently used materials (e.g. $Li_xCoO_2$) typically ranging from $10^{-13}$ cm$^2$/s to $10^{-7}$ cm$^2$/s, it is seen clearly that LFS cannot provide at the moment better kinetics than the state-of-the-art materials. However, by tailoring the particle size of the LFS one can reduce the diffusion coefficient below $10^{-7}$ cm$^2$/s. The result will be published in a forthcoming article.

**Conclusion**

In summary, the relative stability of the LFS has been studied using density-functional total-energy calculations. At ambient condition, LFS stabilizes in the monoclinic ($P2_1$) structure. When an external pressure is applied, monoclinic LFS transforms into orthorhombic ($Pmn2_1$) at the critical load of 0.38 GPa. A further transformation (LFS into $I222$) is then observed at 1.9 GPa. The calculated structural data for modifications are in very good agreement with experimental and theoretical reports. At equilibrium, the energy difference between the $P2_1/c$, *Pnma, Pc, Pna*$2_1$ and modified-$Pmn2_1$ modifications is very small and, as a result, depending upon the method of synthesis one can stabilize either of these phases at ambient conditions. The phonon calculations reveal that LFS is dynamically stable in $P2_1/c$, *Pnma, Pc,* and modified-$Pmn2_1$ structures; a larger pressure makes $I222$ modification become stable above the phase transition point. The calculated single crystal elastic constants indicate that *Pnma, Pna*$2_1$, modified-$Pmn2_1$, Pnmb, and $I222$ phases are mechanically stable. The low energy $P2_1/c$ structure is a dynamically stable but mechanically unstable phase. This might explain why it is often experimentally observed as metastable. The diffusion coefficients in LFS at room temperature are calculated, their values ranging from $10^{-54}$ cm$^2$/s up to $10^{-21}$ cm$^2$/s and therefore suboptimal with respect to the current generation of state-of-the-art materials. More work is therefore required in order to improve the Li-ion diffusion in this class of materials.

**Acknowledgment**

The authors gratefully acknowledge the Research Council of Norway (Grant agreement no.: Nano-MILIB, 143732) project for financial support. PV acknowledges the Research Council of Norway for providing the computer time (under the project number NN2875k) at the Norwegian supercomputer.

**Table 1** The calculated equilibrium structural parameters ($a$, $b$ and $c$ are in Å) and bandgap ($E_g$ in eV) values for $Li_2FeSiO_4$ polymorphs.

| Phase | Lattice parameter | | | | $E_g$ (eV) |
|---|---|---|---|---|---|
| | $a$ | $b$ | $c$ | $β(deg)$ | |
| $Pmn2_1$ | 6.3250(6.295)[a] | 5.3842(5.3454)[a] | 4.9731(4.9624)[a] | 90 | 3.2 |
| $P2_1$ | 8.313(8.2290)[b] | 5.084(5.0200)[b] | 8.282(8.2335)[b] | 99.15(99.20) | 3.2 |
| $P2_1/n$ | 8.2320(8.231)[a] | 5.0168 (5.022)[a] | 8.2348(8.232)[a] | 99.18(99.27) | 3.15 |
| $Pna2_1$ | 6.2812(6.276)[c] | 10.9810(10.973)[c] | 5.0158(5.016)[c] | 90 | 3.06 |
| $Pmn2_1$-mod | 6.3743(6.271)[c] | 5.5755(5.485)[c] | 5.1002(5.017)[c] | 90.5(90.5)[c] | 3.07 |
| $Pc$* | 5.1002 | 5.5755 | 8.1311 | 128.4 | |
| $Pmna$ | 6.341(6.2853)[a] | 10.747(10.6592)[a] | 5.100 (5.0367)[a] | 86.89(84.36) | 3.11 |
| $I222$** | 4.7301 | 4.7237 | 5.5346 | 90 | 3.24 |

[a]From ref. [65]; [b] From ref. [9]; [c]From theoretical calculation by Saracibar *et al*.[10] *According to the structural analysis the modified $Pmn2_1$ structure ($Pmn2_1$-mod) can be described in monoclinic (space group *Pc*; space group number 7) structure. **Structural parameters at the phase transition point.

**Table 2** Computational details for the phonon calculation, calculated zero-point energy (ZPE) from the phonon density of states, and dynamical stability (DS) for different $Li_2FeSiO_4$ polymorphs.

| Phase | Super cell size | Num.atoms | ZPE (eV) | DS |
|---|---|---|---|---|
| $Pmn2_1$ | 2×2×2 | 128 | 0.629 | Stable |
| $P2_1$ | 1×2×1 | 64 | 0.626 | Stable |
| $P2_1/n$ | 1×2×1 | 64 | 0.644 | Stable |
| $Pna2_1$ | 2×1×2 | 128 | 0.665 | Unstable |
| $Pmnb$ | 2×1×2 | 128 | 0.627 | Stable |
| $Pc$* | 2×2×2 | 128 | 0.581 | Stable |
| $I222$** | 3×3×3 | 216 | 0.730 | Stable |

*According to the structural analysis the modified $Pmn2_1$ structure ($Pmn2_1$-mod) can be described in monoclinic (space group *Pc*; space group number 7) structure. **Stable at high pressure.

**Table 3** Calculated barrier height (in eV), Li-Li migration distance (in Å), and diffusion coefficient (D; cm$^2$ S$^{-1}$) for different Li$_2$FeSiO$_4$ polymorphs.

| Phase | Barrier height | | Li-Li distance | diffusion coefficient | Ref. |
|---|---|---|---|---|---|
| | Pathway A | Pathway B | | | |
| *Pmn21* | 0.86[a], 0.87[b], 0.85[c] 0.79[c1] | 0.98[a], 0.97[b], 1.0[c], | 3.7 - 3.74 | 10$^{-14}$/10$^{-13}$ | [a]present study, [b][61], [c][60] [c1][41] |
| *Pbn21* | 0.80[a], 0.82[b], | 0.71[a], 0.72[b], | 3.01, 3.02, 4.24 | 10$^{-14}$/10$^{-12}$ | |
| *P2$_1$* | 0.83[a], 0.84[d], 0.83[e], 0.70[c1] | 0.93[a], 0.95[d], 0.87[e] | 3.7 – 3.9 4.236, 3.537 | 10$^{-13}$/10$^{-11}$ | [d] [42] [e] [63] |
| *Pmn21-cycl* | 0.91[f] | | 3.02, 3.23 | | [f] [59] |
| *P21/n-cycl* | 0.83/0.74/0.60[g] | | 3.07, 3.56 | 10$^{-14}$/10$^{-12}$/10$^{-10}$ | [g] [62] |
| *I222* | 1.23[a] | 1.64[a] | 3.53, 4.53 | 10$^{-22}$/10$^{-54}$ | [a]present study |

**Table 4** The calculated single crystal elastic constants $C_{ij}$ (in GPa), bulk modulus $B$ (in GPa), shear modulus $G$ (in GPa), Possion's ratio $\nu$, Young's modulus $E$ (in GPa), and compressibility (GPa$^{-1}$) for Li$_2$FeSiO$_4$ polymorphs. Subscript $V$ indicates the Voigt bound, $R$ indicates the Reuss bound and $H$ indicates the Hill average.

| Properties | | | | Phase | | | | |
|---|---|---|---|---|---|---|---|---|
| | *Pmn2$_1$* | *P2$_1$* | *P2$_1$/n* | *Pna2$_1$* | *Pc* | *Pc\** | *Pmnb* | *I222* |
| $C_{ij}$ | $C_{11}=218$ | $C_{11}=149$ | $C_{11}=145$ | $C_{11}=119$ | $C_{11}=128$ | $C_{11}=127$ | $C_{11}=206$ | $C_{11}=85$ |
| | $C_{12}=58$ | $C_{12}=50$ | $C_{12}=44$ | $C_{12}=52$ | $C_{12}=53$ | $C_{12}=54$ | $C_{12}=61$ | $C_{12}=168$ |
| | $C_{13}=58$ | $C_{13}=81$ | $C_{13}=72$ | $C_{13}=46$ | $C_{13}=47$ | $C_{13}=56$ | $C_{13}=41$ | $C_{13}=169$ |
| | $C_{22}=135$ | $C_{15}=-22$ | $C_{15}=-14$ | $C_{22}=146$ | $C_{15}=-5$ | $C_{15}=0.3$ | $C_{22}=121$ | $C_{22}=159$ |
| | $C_{23}=46$ | $C_{22}=125$ | $C_{22}=109$ | $C_{23}=39$ | $C_{22}=153$ | $C_{22}=170$ | $C_{23}=40$ | $C_{23}=-25$ |
| | $C_{33}=134$ | $C_{23}=48$ | $C_{23}=38$ | $C_{33}=118$ | $C_{23}=54$ | $C_{23}=59$ | $C_{33}=98$ | $C_{33}=164$ |
| | $C_{44}=35$ | $C_{25}=-5$ | $C_{25}=-0.2$ | $C_{44}=38$ | $C_{25}=-9$ | $C_{25}=1.8$ | $C_{44}=40$ | $C_{44}=28$ |
| | $C_{55}=41$ | $C_{33}=170$ | $C_{33}=146$ | $C_{55}=42$ | $C_{33}=121$ | $C_{33}=135$ | $C_{55}=53$ | $C_{55}=63$ |
| | $C_{66}=42$ | $C_{35}=-25$ | $C_{35}=-22$ | $C_{66}=36$ | $C_{35}=-4$ | $C_{35}=-7$ | $C_{66}=40$ | $C_{66}=58$ |
| | | $C_{44}=41$ | $C_{44}=46$ | | $C_{44}=34$ | $C_{44}=41$ | | |
| | | $C_{46}=-5$ | $C_{46}=-4$ | | $C_{46}=-4$ | $C_{46}=2.9$ | | |
| | | $C_{55}=55$ | $C_{55}=50$ | | $C_{55}=38$ | $C_{55}=40$ | | |
| | | $C_{66}=48$ | $C_{66}=39$ | | $C_{66}=39$ | $C_{66}=42$ | | |
| $B_V$ | 90 | 89 | 78 | 90 | 79 | 85 | 79 | 115 |
| $B_R$ | 85 | 82 | 73 | 84 | 78 | 84 | 71 | 123 |
| $B_H$ | 87 | 86 | 76 | 87 | 78 | 85 | 75 | 119 |
| $G_V$ | 46 | 46 | 44 | 46 | 39 | 42 | 46 | 36 |
| $G_R$ | 44 | 44 | 41 | 43 | 38 | 41 | 43 | 78 |
| $G_H$ | 45 | 45 | 43 | 45 | 38 | 42 | 44 | 57 |
| $\nu_{xy}$ | 0.32 | 0.25 | 0.26 | 0.32 | 0.24 | 0.20 | 0.42 | 1.25 |
| $\nu_{yx}$ | 0.20 | 0.26 | 0.23 | 0.20 | 0.24 | 0.28 | 0.23 | -2.19 |
| $\nu_{zx}$ | 0.2 | 0.44 | 0.41 | 0.2 | 0.26 | 0.34 | 0.12 | -2.10 |
| $\nu_{xz}$ | 0.33 | 0.38 | 0.41 | 0.33 | 0.29 | 0.33 | 0.25 | 1.22 |
| $\nu_{yz}$ | 0.25 | 0.17 | 0.16 | 0.25 | 0.35 | 0.33 | 0.31 | 2.09 |
| $\nu_{zy}$ | 0.25 | 0.20 | 0.18 | 0.25 | 0.28 | 0.24 | 0.27 | 2.07 |
| $E_x$ | 180 | 101 | 103 | 91 | 101 | 98 | 171 | -331 |
| $E_y$ | 112 | 104 | 92 | 120 | 117 | 135 | 95 | 580 |
| $E_z$ | 110 | 118 | 102 | 97 | 92 | 100 | 83 | 572 |
| Compressibility | 0.012 | 0.012 | 0.014 | 0.014 | 0.013 | 0.012 | 0.014 | 0.008 |

*According to the structural analysis the modified *Pmn2$_1$* structure (*Pmn2$_1$-mod*) can be described in monoclinic (space group *Pc*; space group number 7) structure.

# References

[1]     A. K. Padhi, K. S. Nanjundaswamy, and J. B. Goodenough, J. Electrochem. Soc. **144**, 1188 (1997).
[2]     A. Manthiram and J. B. Goodenough, J. Solid State Chem. **71**, 349 (1987).
[3]     A. Manthiram and J. B. Goodenough, J. Power Sources **26**, 403 (1989).
[4]     K. S. Nyten A, Haggstrom L, Gustafsson T, Thomas JO and  J Mater Chem **16**, 2266 (2006).
[5]     Zhong G, Li Y, Yan P, Liu Z, Xie M, and L. H, J. Phys. Chem. C  **114**, 3693 (2010).
[6]     R. Dominko, M. Bele, M. Gaberscek, A. Meden, M. Remskar, and J. Jamnik, Electrochem. Comm. **8**, 217 (2006).
[7]     M. Arroyo-de Dompablo, R. Dominko, J. Gallardo-Amores, L. Dupont, G. Mali, H. Ehrenberg, J. Jamnik, and E. Moran, Chem. Mater. **20**, 5574 (2008).
[8]     D. M. Arroyo-de, J. Gallardo-Amores, J. Garc´ıa-Mart´ınez, E. Mor´an, J. M. Tarascon, and M. Armand, Solid State Ionics **179**, 1758 (2008).
[9]     S. Nishimura, S. Hayase, R. Kanno, M. Yashima, N. Nakayama, and A. Yamada, J. Am. Chem. Soc. **130**, 13212 (2008).
[10]    A. Saracibar, A. Van der Ven, and M. E. Arroyo-de Dompablo, Chem.f Mater. **24**, 495 (2012).
[11]    A. R. Armstrong, N. Kuganathan, M. S. Islam, and P. G. Bruce, J. Am. Chem. Soc. **133**, 13031 (2011).
[12]    A. Liivat and O. Thomas J, Solid State Ionics **192**, 58 ( 2011).
[13]    G. Mali, C. Sirisopanaporn, C. Masquelier, D. Hanzel, and R. Dominko, Chem. Mater. **23**, 2735 (2011).
[14]    C. Sirisopanaporn, A. Boulineau, R. Dominko, D. Hanzel, A. R. Armstrong, P. Bruce, and C. Masquelier, Inorg. Chem. **49**, 7446 (2010).
[15]    A. Nyten, S. Kamali, L. Haggstrom, T. Gustafsson, and J. O. Thomas, J. Mater.s Chem. **16**, 2266 (2006).
[16]    G. Kresse and J. Furthmüller, Phys. Rev. B **54**, 11169 (1996).
[17]    G. Kresse and J. Furthmüller, Comput. Mater. Sci. **6**, 15 (1996).
[18]    J. P. Perdew, K. Burke, and M. Ernzerhof, Phys. Rev. Lett. **77**, 3865 (1996).
[19]    A. I. Liechtenstein, V. I. Anisimov, and J. Zaanen, Phys. Rev. B **52**, R5467 (1995).
[20]    S. L. Dudarev, G. A. Botton, S. Y. Savrasov, Z. Szotek, W. M. Temmerman, and A. P. Sutton, Phys. Status Solidi A **166**, 429 (1998).
[21]    P. Vinet, J. H. Rose, J. Ferrante, and J. R. Smith, J. Phys.: Condens. Matter **1**, 1941 (1989).
[22]    A. Togo, F. Oba, and I. Tanaka, Phys. Rev. B **78**, 134106 (2008).
[23]    H. J. Monkhorst and J. D. Pack, Phys. Rev. B **13**, 5188 (1976).
[24]    H. Jonsson, G. Mills, and K. W. Jacobsen, in *Classical and Quantum Dynamics in Condensed Phase Simulations* (WORLD SCIENTIFIC, 2011), pp. 385.
[25]    G. Henkelman and H. Jónsson, J. Chem. Phys. **113**, 9978 (2000).
[26]    P. Tarte and R. Cahay, C. R. Acad. Sci. **271**, 777 (1970 ).
[27]    G. Quoirin, F. Taulelle, L. Dupont, and C. Masquelier, 211th ECS Meeting **Abstract no. 98** (2007).
[28]    A. Nyten, A. Abouimrane, M. Armand, T. Gustafsson, and J. O. Thomas, Electrochem. Commun.  **7**, 156 (2005).
[29]    C. Frayret, C. Masquelier, A. Villesuzanne, M. Morcrette, and J. M. Tarascon, Chem. Mater. **21** 1861 (2009).
[30]    L. D. Iskhakova and V. B. Rybakov, Crystallogr. Rep. **48,**, 44 (2003).

[31]  H. Yamaguchi and K. Akatsuka, Acta Crystallogr. **B35**, 2678 (1979).
[32]  C. Lyness, B. Delobel, A. R. Armstrong, and P. G. Bruce, Chem. Commun. **46**, 4890 (2007).
[33]  V. V. Politaev, A. A. Petrenko, V. B. Nalbandyan, B. S. Medvedev, and E. S. Shvetsova, J. Solid State Chem. **180**, 1045 (2007).
[34]  C. Eames, A. R. Armstrong, P. G. Bruce, and M. S. Islam, Chem. Mater. **24**, 2155 (2012).
[35]  N. G. Hörmann and A. Groß, J. Solid State Electrochem. (2013).
[36]  *Inorganic Crystal Structure Database (ICSD) Version 2011/2*.
[37]  P. Vajeeston, P. Ravindran, A. Kjekshus, and H. Fjellvåg, J. Alloys Compd. **363**, L8 (2004).
[38]  P. Vajeeston, P. Ravindran, B. C. Hauback, and H. Fjellvag, Int. J. Hydrogen Energy **36**, 10149 (2011).
[39]  P. Vajeeston, P. Ravindran, A. Kjekshus, and H. Fjellvåg, Physical Review B **71**, 092103 (2005).
[40]  X. Lu, H.-C. Chiu, Z. Arthur, J. Zhou, J. Wang, N. Chen, D.-T. Jiang, K. Zaghib, and G. P. Demopoulos, J. Power Sources **329**, 355 (2016).
[41]  X. Lu, H.-C. Chiu, K. H. Bevan, D.-T. Jiang, K. Zaghib, and G. P. Demopoulos, J. Power Sources **318**, 136 (2016).
[42]  R. B. Araujo, R. H. Scheicher, J. S. de Almeida, A. Ferreira da Silva, and R. Ahuja, Solid State Ionics **247–248**, 8 (2013).
[43]  V. B. Deyirmenjian, V. Heine, M. C. Payne, V. Milman, R. M. Lynden-Bell, and M. W. Finnis, Phys. Rev. B **52**, 15191 (1995).
[44]  M. Marlo and V. Milman, Phys. Rev. B **62**, 2899 (2000).
[45]  V. Milman and M. C. Warren, J. Phys.: Condens. Matter **13**, 5585 (2001).
[46]  P. Ravindran, P. Vajeeston, R. Vidya, A. Kjekshus, and H. Fjellvåg, Phys. Rev. B **64**, 224509 (2001).
[47]  A. de Vita, I. Manassidis, J. S. Lin, and M. J. Gillan, Europhys. Lett. **19**, 605 (1992).
[48]  P. Ravindran, L. Fast, P. A. Korzhavyi, B. Johansson, J. Wills, and O. Eriksson, J. Appl. Phys. **84**, 4891 (1998).
[49]  B. B. Karki, S. J. Clark, M. C. Warren, H. C. Hsueh, G. J. Ackland, and J. Crain, J. Phys.: Cond. Matt. **9**, 375 (1997).
[50]  F.I. Fedorov, *Theory of Elastic Waves in Crystals* (Plenum, New York, 1968).
[51]  J. F. Nye, *Physical Properties of Crystals* (Oxford: Oxford University Press, 1985).
[52]  J. P. Watt, J. Appl. Phys. **51** 1520 (1980).
[53]  A. Reuss, ZAMM - Journal of Applied Mathematics and Mechanics / Zeitschrift für Angewandte Mathematik und Mechanik **9**, 49 (1929).
[54]  R. Hill, Proc. Phys. Soc., London, **65**, 350 (1952).
[55]  W. Voigt, *Lehrbuch der Kristallphysik* Leipzig, 1928).
[56]  S. F. Pugh, Philos. Mag. **45**, 823 (1954).
[57]  S. Zhang., C. Deng., and S. Yang., Electrochem Solid-State Let. **12**, A136 (2009).
[58]  K. Kang, D. Morgan, and G. Ceder, Phys. Rev. B **79**, 014305 (2009).
[59]  A. R. Armstrong, N. Kuganathan, M. S. Islam, and P. G. Bruce, J. Am. Chem.l Soc. **133**, 13031 (2011).
[60]  R. B. Araujo, R. H. Scheicher, J. S. de Almeida, A. Ferreira da Silva, and R. Ahuja, Solid State Commun. **173**, 9 (2013).
[61]  A. Liivat and J. O. Thomas, Solid State Ionics **192**, 58 (2011).
[62]  P. Zhang, Y. Zheng, S. Yu, S. Q. Wu, Y. H. Wen, Z. Z. Zhu, and Y. Yang, Electrochimica Acta. **111**, 172 (2013).
[63]  D. Su, H. Ahn, and G. Wang, Appl. Phys. Lett. **99**, 141909 (2011).

[64]   Y. Sun, X. Lu, R. Xiao, H. Li, and X. Huang, Chem. Mater. **24**, 4693 (2012).
[65]   C. Sirisopanaporn, C. Masquelier, P. G. Bruce, A. R. Armstrong, and R. Dominko, J. Am. Chem. Soc. **133**, 1263 (2010).

# Supplemental Information for Article Titled: "First-principles study of structural stability, dynamical and mechanical properties of Li$_2$FeSiO$_4$ polymorphs"

P. Vajeeston,[*] and H. Fjellvåg

*Center for Materials Sciences and Nanotechnology, Department of Chemistry, University of Oslo, P.O. Box 1033 Blindern, N-0315 Oslo, Norway*

E-mail: ponniahv@kjemi.uio.no

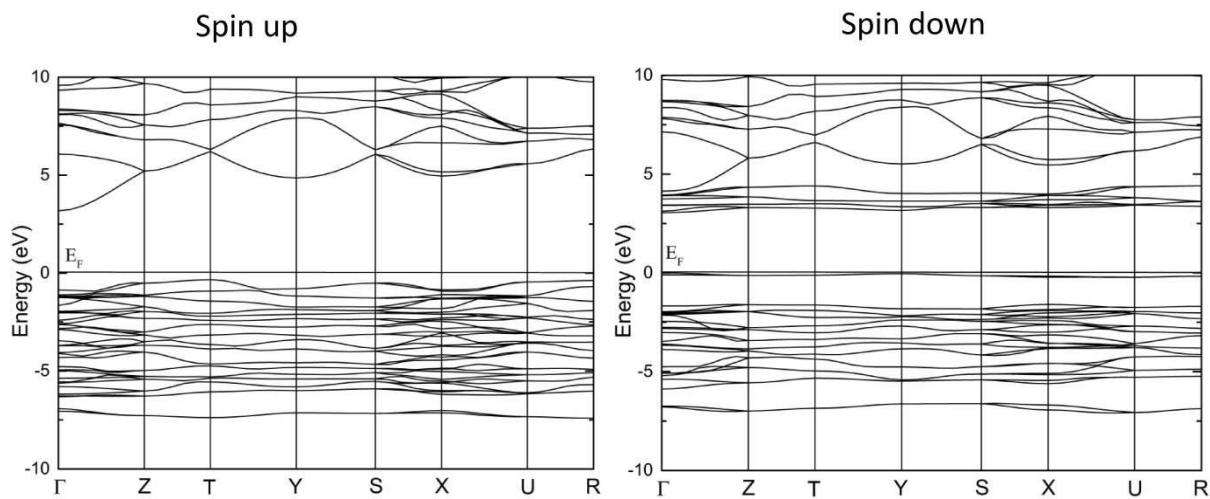

**Fig.S1** Calculated electronic up-spin and down-spin band structure of Li$_2$FeSiO$_4$ in *Pmn2$_1$* structure. The Fermi level is set to zero.

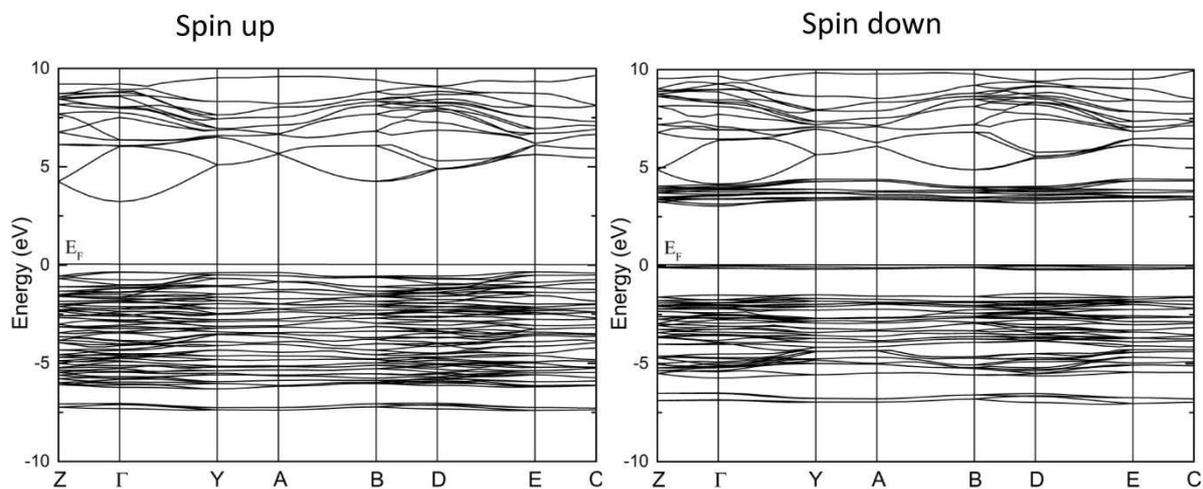

**Fig.S2** Calculated electronic up-spin and down-spin band structure of Li$_2$FeSiO$_4$ in *P2$_1$* structure. The Fermi level is set to zero.

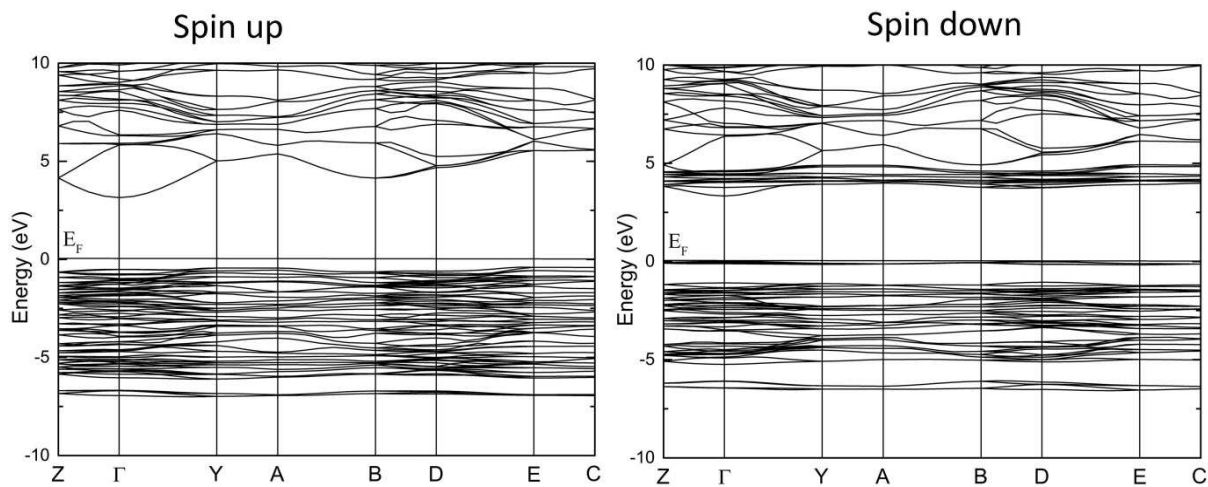

**Fig.S3** Calculated electronic up-spin and down-spin band structure of Li$_2$FeSiO$_4$ in *P2$_1$/n* structure. The Fermi level is set to zero.

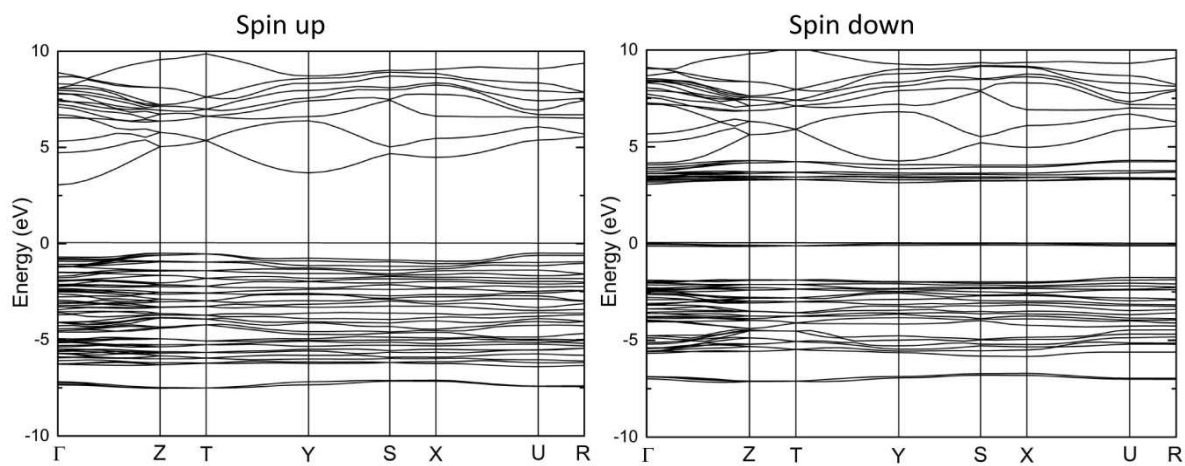

**Fig.S4** Calculated electronic up-spin and down-spin band structure of Li$_2$FeSiO$_4$ in *Pbn2$_1$* structure. The Fermi level is set to zero.

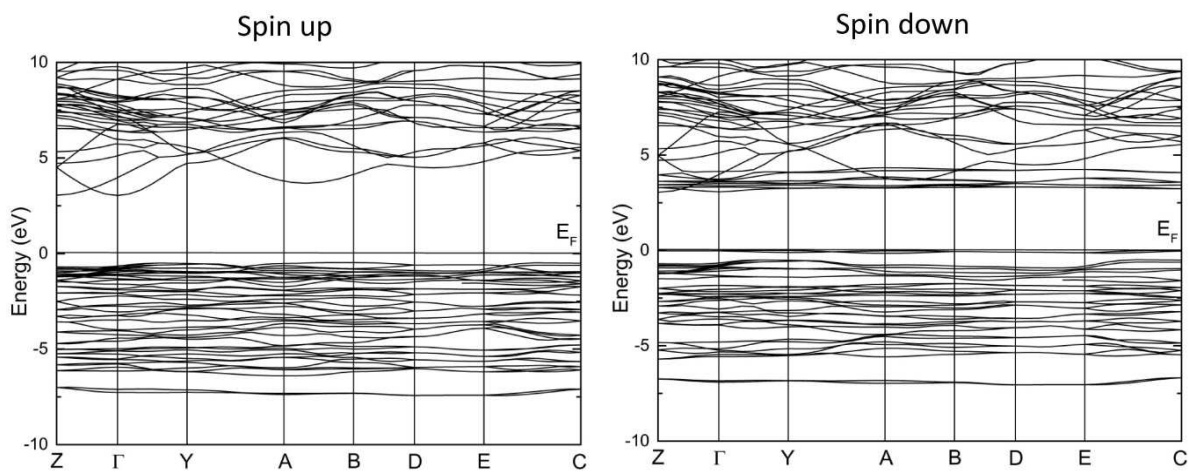

**Fig.S5** Calculated electronic up-spin and down-spin band structure of Li$_2$FeSiO$_4$ in *Pmn2$_1$-mod* structure. According to the structural analysis the modified *Pmn2$_1$* structure (*Pmn2$_1$-mod*) can be described in monoclinic (space group *Pc*; space group number 7) structure. The Fermi level is set to zero.

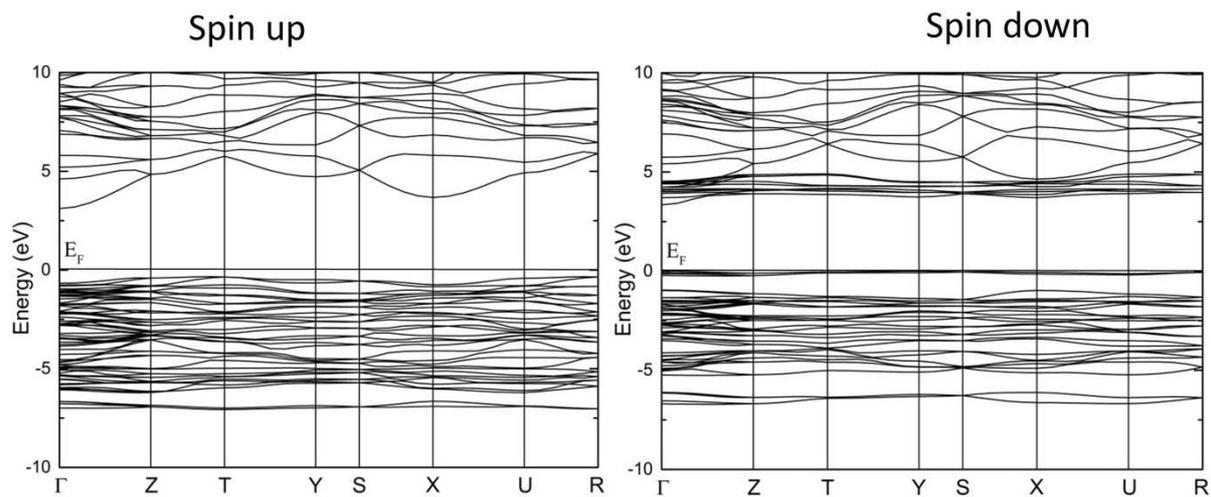

**Fig.S6** Calculated electronic up-spin and down-spin band structure of Li$_2$FeSiO$_4$ in *Pmnb* structure. The Fermi level is set to zero.

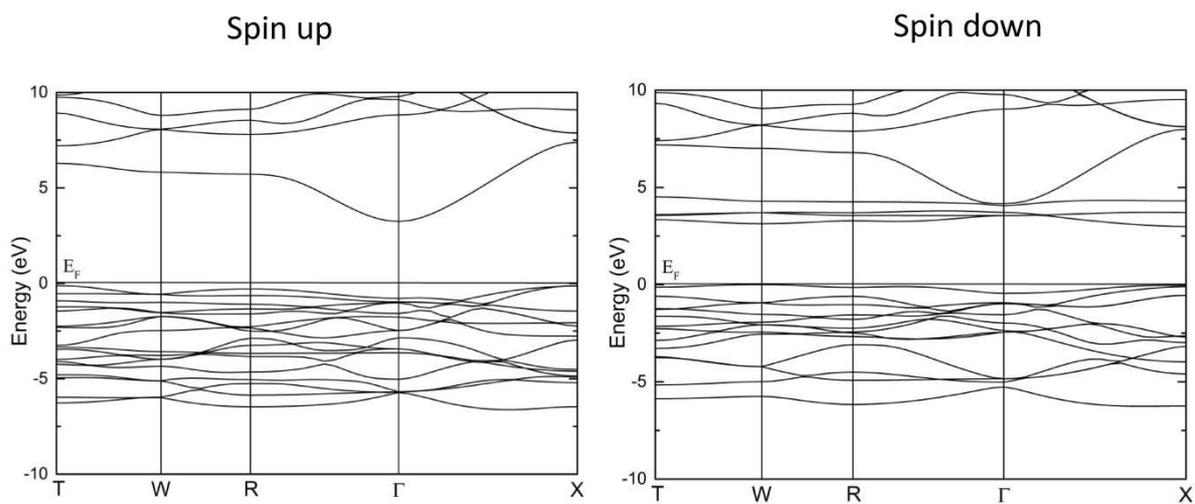

**Fig. S7** Calculated electronic up-spin and down-spin band structure of Li$_2$FeSiO$_4$ in *I222* structure. The Fermi level is set to zero.

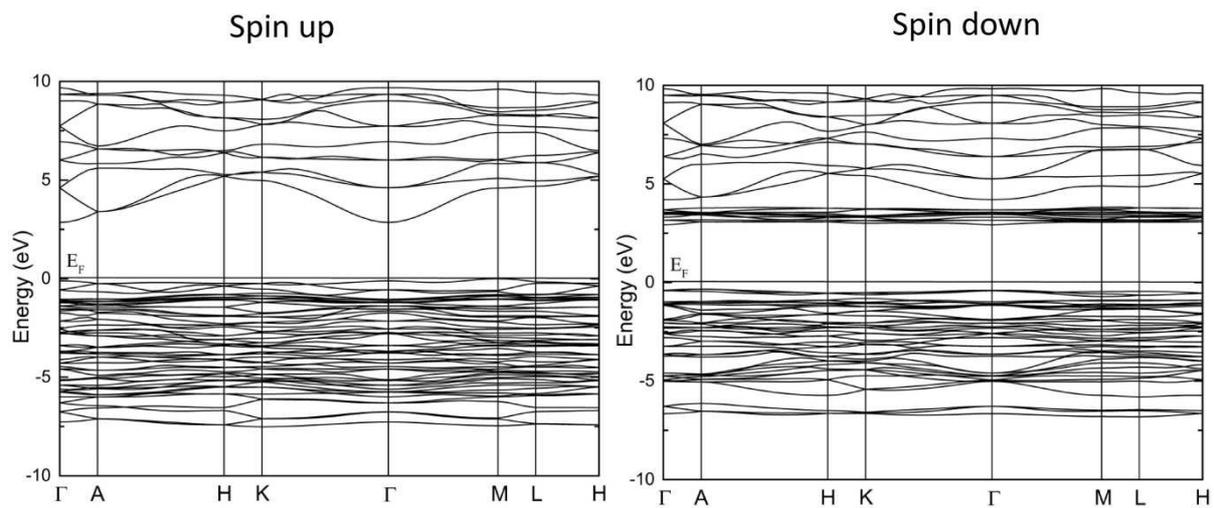

**Fig.S8** Calculated electronic up-spin and down-spin band structure of Li$_2$FeSiO$_4$ in P3121 structure. The Fermi level is set to zero.

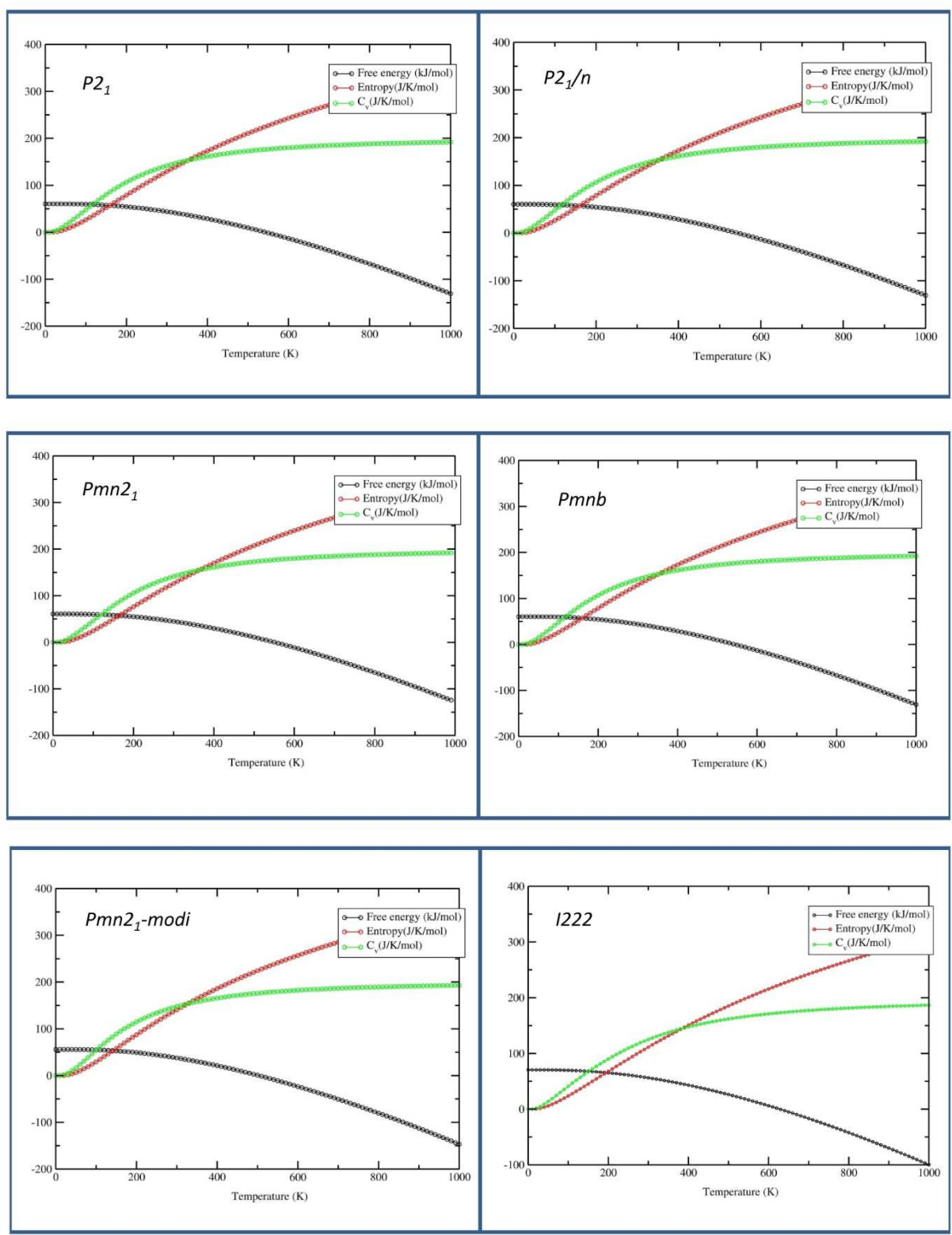

**Fig.S9** Calculated free energy (in kJ/mol), entropy (in J/K/mol.) and lattice heat capacity (Cv; in J/K/mol.) as a function of temeperature for $Li_2FeSiO_4$ in *P21, P21/n, Pmn21, Pmnb, Pmn21-modi* and *I222* modifications.